\def\papertitle{A Comparison of Virtual Analog Modelling Techniques for Desktop and Embedded Implementations}
\def\paperauthorA{Jatin Chowdhury}

\documentclass[twoside,a4paper]{article}
\usepackage{dafx_19}
\usepackage{amsmath,amssymb,amsfonts,amsthm}
\usepackage{euscript}
\usepackage[latin1]{inputenc}
\usepackage[T1]{fontenc}
\usepackage{ifpdf}

\usepackage[english]{babel}
\usepackage{caption}
\usepackage{subfig} 
\usepackage{xcolor}

\ninept

\usepackage{times}

\newif\ifpdf
\ifx\pdfoutput\relax
\else
   \ifcase\pdfoutput
      \pdffalse
   \else
      \pdftrue
\fi

\ifpdf 
  \usepackage[pdftex,
    pdftitle={\papertitle},
    pdfauthor={\paperauthorA},
    colorlinks=false, 
    bookmarksnumbered, 
    pdfstartview=XYZ 
  ]{hyperref}
  \usepackage[pdftex]{graphicx}
  \usepackage[figure,table]{hypcap}
\else 
  \usepackage[dvips]{epsfig,graphicx}
  \usepackage[dvips,
    colorlinks=false, 
    bookmarksnumbered, 
    pdfstartview=XYZ 
  ]{hyperref}
  \usepackage[figure,table]{hypcap}
\fi

\usepackage{tikz}
\usetikzlibrary{positioning}
\usepackage[american]{circuitikz}
\usepackage{tkz-euclide}
\usepackage{cleveref}
\usepackage{siunitx}

\usepackage{listings}
\definecolor{codegreen}{rgb}{0,0.6,0}
\definecolor{codegray}{rgb}{0.5,0.5,0.5}
\definecolor{codepurple}{rgb}{0.58,0,0.82}
\definecolor{backcolour}{rgb}{0.95,0.95,0.92}
 
\lstdefinestyle{mystyle}{
    backgroundcolor=\color{backcolour},   
    commentstyle=\color{codegreen},
    keywordstyle=\color{magenta},
    numberstyle=\tiny\color{codegray},
    stringstyle=\color{codepurple},
    basicstyle=\footnotesize,
    columns=flexible,
    breakatwhitespace=false,         
    breaklines=true,                 
    captionpos=b,                    
    keepspaces=true,                               
    showspaces=false,                
    showstringspaces=false,
    showtabs=false,                  
    tabsize=4
}
 
\DeclareMathAlphabet{\mathpzc}{OT1}{pzc}{m}{it}

\title{\papertitle}

\affiliation{
\paperauthorA \,}
{\href{http://ccrma.stanford.edu}{Center for Computer Research in Music and Acoustics} \\ Stanford University \\ Palo Alto, CA \\ {\tt \href{mailto:jatin@ccrma.stanford.edu}{jatin@ccrma.stanford.edu}}}

\begin{document}
\ifpdf 
  \DeclareGraphicsExtensions{.png,.jpg,.pdf}
\else  
  \DeclareGraphicsExtensions{.eps}
\fi

\graphicspath{{./Figures/}}

\maketitle
\begin{abstract}
We develop a virtual analog model of the Klon Centaur
guitar pedal circuit, comparing various circuit modelling techniques.
The techniques analyzed include traditional modelling techniques such
as nodal analysis and Wave Digital Filters, as well as a machine-learning
technique using recurrent neural networks. We examine these techniques
in the contexts of two use cases: an audio plug-in designed to be
run on a consumer-grade desktop computer, and a guitar pedal-style effect
running on an embedded device. Finally, we discuss the advantages and
disdvantages of each technique for modelling different circuits, and
targeting different platforms.
\end{abstract}

\section{Introduction}
The Klon Centaur is an overdrive guitar pedal designed by Bill
Finnegan in the early 1990's, that has developed cult acclaim
amongst guitarists \cite{Finnegan}. The circuit is notable for
producing ``transparent distortion'' \cite{electrosmash},
a term used to describe the way the pedal seems to add distortion
to a guitar's sound without otherwise affecting the tone. While
the original manufacturing run of the pedal ended in 2004, many
``clones'' of the pedal have been produced by other manufacturers,
adding to its cult following.
\newline\newline
Circuit modelling is typically broken down into ``white-box'' and
``black-box'' approaches \cite{Germain}. A ``white-box'' approach
uses knowledge of the internal mechanisms of the circuit, often
modelling the physical interactions of the electrical components.
Popular white-box methods include nodal analysis \cite{Yeh},
Port-Hamiltonian analysis \cite{PortHamiltonian}, Wave Digital
Filters \cite{Fettweis,KurtThesis}, and nonlinear state space
analysis \cite{StateSpace}.
\newline\newline
``Black-box'' circuit modelling methods generally use measurements
taken from the circuit being modelled and attempt to model the
response of the circuit without knowledge of the internal workings
of the system. Traditional black-box techniques include impulse
response measurements \cite{sasp} and extensions thereof, including
the Weiner-Hammerstein method \cite{Germain}. Recently, researchers
have begun using machine learning methods for black-box modelling.
Damsk\"agg et. al. model several guitar distortion circuits, using a
WaveNet style architecture to generate an output signal sample-by-sample
\cite{WaveNetVA}. Parker et. al. use deep fully-connected networks to
approximate nonlinear state-space solutions for the Korg MS-20 filter
circuit, effectively a ``grey-box'' approach \cite{NLML}. Finally, Wright
et. al. use a recurrent neural network to model the behavior of guitar
distortion circuits with control parameters \cite{VArnn}.
\newline\newline
The structure of the paper will be as follows: in \S2 we give
background information on circuit modelling using nodal analysis
and Wave Digital Filters. \S3 describes the use of recurrent neural
networks for circuit modelling, and outlines the model and training
process used for emulating the ``Gain Stage'' circuit from the Klon
Centaur. In \S4 we discuss the real-time implementation of a complete
emulation of the Klon Centaur pedal using the methods outlined in the
previous sections. \S5 shares the results of Klon Centaur emulation as
well as recommendations for circuit modelling using the methods
discussed here.

\section{Traditional Circuit Modelling Techniques}
First, we examine the use of traditional circuit modelling techniques,
specifically nodal analysis and Wave Digital filters, using
sub-circuits from the Klon Centaur as examples.

\subsection{Nodal Analysis}
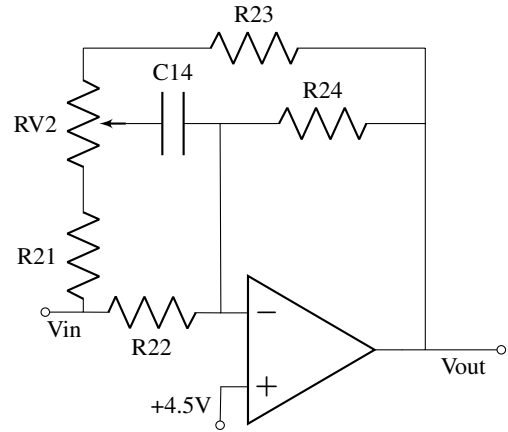
\begin{figure}
    \centering
    \begin{circuitikz} \draw
        (0, 0) node[op amp] (opamp) {}
        (opamp.+) to[short, l_=+4.5V,-o] (-1.2, -1.0)
        (opamp.-) to[R, l=R22] (-3, 0.5)
        to[short, l=Vin,-o] (-3.5, 0.5)
        (-3, 0.5) to[R, l=R21] (-3, 2.0)
        (-3, 4.0) to[american potentiometer, l_=RV2, n=mypot] (-3, 2.0)
        (mypot.wiper) to[C, l=C14] (-1.2, 3.0)
        (opamp.-) -- (-1.2, 3.0)
        to[R, l=R24] (1.5, 3.0)
        (-3, 4.0) to[R, l=R23] (1.5, 4.0)
        -- (1.5, 0.0)
        (opamp.out) -- (1.5, 0.0) to[short, l_=Vout,-o] (2.5, 0.0)
      ;
    \end{circuitikz}
    \caption{\label{fig:ToneControl}{\it Klon Centaur Tone Control Circuit}}
\end{figure}

\begin{figure}
    \centering
    \includegraphics[width=0.5\textwidth]{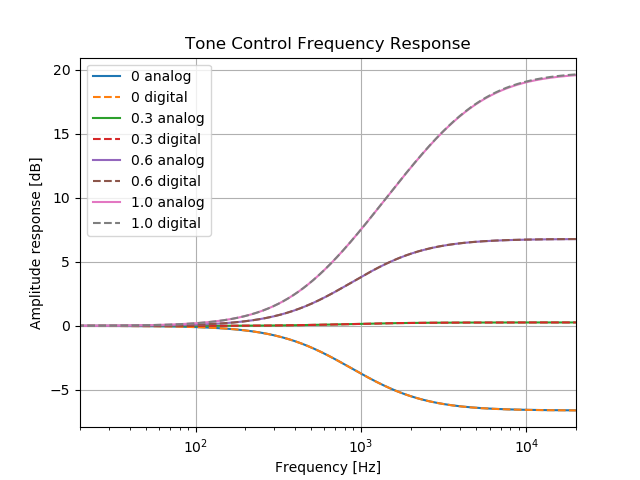}
    \caption{\label{fig:ToneFreq} {\it Tone control frequency response
    at various values of the Treble parameter, comparing the responses
    of the analog filter with the digital model.}}
\end{figure}

The process for creating a digital model of a circuit using nodal
analysis is as follows:
\begin{enumerate}
    \item Convert the circuit into the Laplace domain.
    \item Form a Laplace domain transfer function of the circuit.
    \item Use a conformal map to transform the circuit into the digital domain.
\end{enumerate}
As an example circuit, we examine the Tone Control circuit from the
Klon Centaur (see \cref{fig:ToneControl}).
The first step is to convert the circuit into the Laplace domain, using
the Laplace variable $s = j\omega$. The impedances for each principle
circuit component: resistors ($Z_R$), capacitors ($Z_C$), and inductors
($Z_L$), are as follows:
\begin{equation}
    Z_R = R, \quad Z_C = \frac{1}{Cs}, \quad Z_L = Ls
\end{equation}
From there, using linear circuit theory, one can construct a Laplace
Domain transfer function for the circuit. Note that this assumes an
ideal operational amplifier operating in its linear region. For more
information on this process, see \cite{Maby}. For the tone control
circuit, the Laplace domain transfer function can be written as:
\begin{equation}
    \frac{V_{out}(s)}{V_{in}(s)} = {\scriptscriptstyle \frac{C_{14}\left(\frac{1}{R_{22}} + \frac{1}{R_{21} + R_{v2b}}\right)s
    + \frac{1}{R_{22}}\left(\frac{1}{R_{21} + R_{v2b}} + \frac{1}{R_{23} + R_{v2a}}\right)}{
      C_{14}\left(\frac{1}{R_{23} + R_{v2a}} + \frac{1}{R_{24}}\right)s
    + \frac{-1}{R_{24}}\left(\frac{1}{R_{21} + R_{v2b}} + \frac{1}{R_{23} + R_{v2a}}\right)}}
\end{equation}
Note that we refer to the the section of potentiometer $R_{v2}$ that is
above the wiper as $R_{v2a}$, and the section below as $R_{v2b}$, and that
we ignore the DC offset created by the $4.5$V voltage source at the positive
terminal of the op-amp.
\newline\newline
Next we use a conformal map to transform the transfer function from the
Laplace domain to the z-plane where it can be implemented as a digital
filter. The most commonly used conformal map is the bilinear transform,
defined as
\begin{equation}
    s \leftarrow \frac{2}{T} \frac{1 - z^{-1}}{1 + z^{-1}}
\end{equation}
Where $T$ is the sample period of the digital system. For more
information on the use of the bilinear transform to digitize an
analog system, see \cite{pasp}. The resulting filter is known as a
``high-shelf'' filter, that accentuates high frequency content in
the signal. The resulting frequency response of the digital model,
validated against the response of the analog circuit is shown in
\cref{fig:ToneFreq}.

\subsubsection{Advantages and Limitations}
The advantages of nodal analysis are that the circuit model
is simple and computationally efficient. The model can be constructed
with minimal knowledge of circuit theory, and a basic understanding
of digitial signal processing. The main disdvantage is that nodal analysis
cannot be used for nonlinear circuits, though it can be extended
to model this class of circuits through modified nodal analysis (MNA)
\cite{MNA}. Another disadvantage of nodal analysis-based methods
is that (typically) large portions of the system need to be recomputed
when a circuit element is changed, such as a potentiometer. While this
computation is fairly simple in the example shown here, it can become
vastly more difficult for more complex systems.

\subsection{Wave Digital Filters}
\begin{figure}
    \centering
    \includegraphics[width=0.5\textwidth]{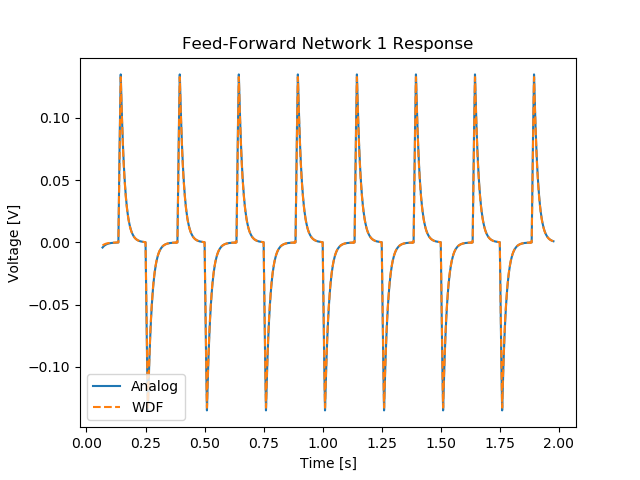}
    \caption{\label{fig:wdfresults}{\it WDF simulation results compared to the analog reference.}}
\end{figure}

The Wave Digital Filter (WDF) formalism allows circuits to be modelled in
modular and flexible manner. Originally developed by Alfred Fettweis
in the 1970's \cite{Fettweis}, WDFs have recently gained popularity in
modelling audio circuits, and have been extended to model a wider class
of circuits \cite{KurtThesis}. The WDF formalism defines each circuit element
as a port with some characteristic resistance $R_0$, and uses wave variables
passing through each port, rather than the typical voltage and current variables.
The incident wave at a certain port is defined as:
\begin{equation}
    a = v + R_0 i
\end{equation}
where $v$ is the voltage across the port, and $i$ is the current passing
through the port. The reflected wave is similarly defined as:
\begin{equation}
    b = v - R_0 i
\end{equation}
A Wave Digital Filter defines circuit elements (resistors, capacitors,
inductors, etc.) in the wave domain, and allows the elements to be connected
by series and parallel adaptors also defined in the wave domain. The
full derivation of these WDF elements is given in \cite{Fettweis} and
\cite{KurtThesis}. 
\newline\newline
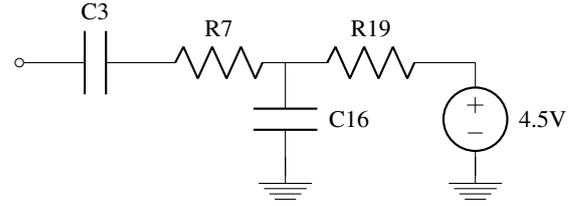
\begin{figure}
    \centering
    \begin{circuitikz} \draw
        (0, 0) node[short, l=Vin] {}
        to[C, l=C3, o-] (2, 0)
        to[R, l=R7]     (3.25, 0)
        to[R, l=R19]    (6, 0)
        to[V, l=4.5V]   (6, -1.5)
        (3.5, 0) to[C, l=C16] (3.5, -1.5)
        (3.5, -1.25) node[ground]() {}
        (6.0, -1.25) node[ground]() {}
      ;
    \end{circuitikz}
    \caption{\label{fig:ff1}{\it Klon Centaur Feed-Forward Network 1 Circuit}}
\end{figure}
\begin{figure}
    \centering
    \begin{tikzpicture}[node distance=1.25cm]
        \tikzset{
            arrow/.style = {thick,->,>=stealth}
        }
        \node (Vin) {$V_{in}$};
        \node (S1)  [right of=Vin] {$\mathcal{S}_1$};
        \node (C3)  [right of=S1, below of=S1, yshift= 0.5cm] {$C_3$};
        \node (S2)  [right of=S1, above of=S1, yshift=-0.5cm] {$\mathcal{S}_2$};
        \node (R7)  [right of=S2, above of=S2, yshift=-0.5cm] {$R_7$};
        \node (P1)  [right of=S2, below of=S2, yshift= 0.5cm] {$\mathcal{P}_1$};
        \node (C16) [right of=P1, below of=P1, yshift= 0.5cm] {$C_{16}$};
        \node (S3)  [right of=P1, above of=P1, yshift=-0.5cm] {$\mathcal{S}_3$};
        \node (R19) [right of=S3, above of=S3, yshift=-0.5cm] {$R_{19}$};
        \node (V45) [right of=S3, below of=S3, yshift= 0.5cm] {$V_{4.5}$};

        \draw [arrow] (Vin) -- (S1);
        \draw [arrow] (S1) -- (C3);
        \draw [arrow] (S1) -- (S2);
        \draw [arrow] (S2) -- (R7);
        \draw [arrow] (S2) -- (P1);
        \draw [arrow] (P1) -- (C16);
        \draw [arrow] (P1) -- (S3);
        \draw [arrow] (S3) -- (R19);
        \draw [arrow] (S3) -- (V45);
    \end{tikzpicture}
    \caption{\label{fig:wdftree}{\it WDF tree for the Klon Centaur Feed-Forward Network 1 Circuit.
    $\mathcal{S}$ and $\mathcal{P}$ nodes refer to series and parallel
    adaptors respectively.}}
\end{figure}
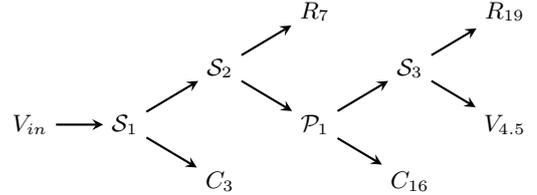
Once each circuit element and adaptor has been defined, they are connected
together in a structure often referred to as a WDF tree. As an example,
we examine the ``feed-forward network 1'' from the Klon Centaur circuit
(see \cref{fig:ff1}). The corresponding WDF tree is shown in \cref{fig:wdftree}.
Simulation results compared to the analog reference are shown
in \cref{fig:wdfresults}.

\subsubsection{Advantages and Limitations}
The primary advantage of the Wave Digital approach is its modularity.
The ability to construct a circuit model with each circuit component
treated completely independently in the digital domain opens up many
interesting possibilities for circuit prototyping, modelling circuit-bent
instruments, and more. Additionally, this modularity allows each circuit
component to be discretized separately, even using different conformal
maps, which can improve model behavior for certain classes of circuits
(see \cite{Germain}). Finally, the separability of components means that
when a component is changed (e.g. a potentiometer), the component change
is propagated so that only components with behavior that depends on the
impedance of the changed component need to be recomputed.
\newline\newline
The main disdvantage of WDFs is their difficulty in handling circuits with
complex topologies or multiple nonlinearities. While the recent addition
of $\mathcal{R}$-type adaptors to the Wave Digital formalism \cite{KurtThesis}
has begun to make these circuits tractable, the WDF models of these types of
circuits are significantly more computationally complex. Further, the use of
$\mathcal{R}$-type adaptors can somewhat compromise the modularity that makes
WDFs advantageous in the first place.

\section{Recurrent Neural Network Model}
While several styles of machine-learning based models have been used for
modelling analog audio circuitry \cite{WaveNetVA,NLML,MartinezReissDNN},
we choose the recurrent neural network approach developed in \cite{VArnn}
as our starting point. Using a recurrent neural network (RNN) allows the for a
significantly smaller neural network than would be possible with a traditional
deep neural network or convolutional neural network, meaning that the network
can be evaluated much faster for real-time use, while maintaining a smaller
memory footprint (an important advantage on embedded platforms). Additionally,
recurrent neural networks are a sensible candidate for modelling distortion
circuits, particularly circuits with stateful behavior, given the fact that
recurrent network building blocks, such as gated recurrent units, themselves
resemble audio distortion effects and can directly be used as such
\cite{chowdhury:complexNL:2020}.
\newline\newline
In the following paragraphs, we outline the use of an RNN for modelling
the gain stage circuit from the Klon Centaur pedal. While our model
is similar to the model used in \cite{VArnn}, it differs in some
notable ways. For instance, the model described in \cite{VArnn} accepts
the values of control parameters to the circuit as inputs to the RNN,
however we were unable to successfully train a network in this fashion.
Instead, we construct separate networks for five different values of the
``Gain'' parameter, and fade between the outputs of the networks in
real-time in the final implementation of the model. Other differences are
outlined further below.

\subsection{Model Architecture}
\begin{figure}
    \centering
    \begin{tikzpicture}[node distance=1.5cm]
        \tikzset{
            mynode/.style = {rectangle, rounded corners, line width=0.8pt, minimum width=2cm, minimum height=0.75cm, text centered, draw=black, fill=white},
            arrow/.style = {thick,->,>=stealth}
        }
        \node (input) {Input $x[n]$};
        \node (rlayer) [mynode, right of=input, xshift=1.25cm] {Recurrent Layer};
        \coordinate[right of=rlayer, xshift=0.5cm] (h0) ;
        \node (h0Label) [right of=h0, xshift=-0.15cm] {Current State $h[n]$};
        \node (z1) [rectangle, draw=black, below of=h0, yshift=0.5cm] {$z^{-1}$};
        \coordinate[below of=rlayer, yshift=0.5cm] (h1);
        \node (h1Label) [below of=h1, yshift=1.15cm] {Previous State $h[n-1]$};

        \draw [arrow] (input) -- (rlayer);
        \draw [thick] (rlayer) -- (h0);
        \draw [arrow] (h0) -- (z1);
        \draw [thick] (z1) -- (h1);
        \draw [arrow] (h1) -- (rlayer);

        \node (dense) [mynode, above of=rlayer] {Fully Connected Layer};
        \node (out) [above of=dense] {Output $y[n]$};
        \draw [arrow] (rlayer) -- (dense);
        \draw [arrow] (dense) -- (out);

    \end{tikzpicture}
    \caption{\label{fig:rnn_arch}{\it RNN Architecture.}}
\end{figure}
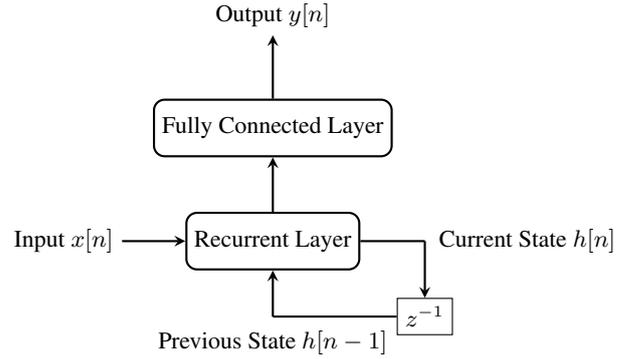

The model architecture described in \cite{VArnn} consists of a single recurrent
layer followed by a fully connected layer consisting of a single ``neuron''
(see \cref{fig:rnn_arch}). In our models, we use a recurrent layer made up
of 8 Gated Recurrent Units. For training, all models are implemented in
\texttt{Python} using the \texttt{Keras} framework \cite{chollet2015keras}.
\subsubsection{Recurrent Layer}
Recurrent layers are typically comprised of one of two types of
recurrent units: Long Short-Term Memory units (LSTMs) or Gated
Recurrent Units (GRUs). For this application, we choose to use
GRUs \cite{gru_original} since they require fewer operations,
allowing for faster computation, and since they requre fewer weights,
thereby allowing the model to have a smaller memory footprint. The
GRU consists of three ``gates'': the update gate $z[n]$, reset gate
$r[n]$, and the new gate $c[n]$. These gates are used to compute the
cell's current output $h[n]$ from its current input $x[n]$ and previous
output $h[n-1]$ as follows:
\begin{equation}
    z[n] = \sigma(W_z x[n] + U_z h[n-1] + b_z)
\end{equation}
\begin{equation}
    r[n] = \sigma(W_r x[n] + U_r h[n-1] + b_r)
\end{equation}
\begin{equation}
    c[n] = \tanh(W_c x[n] + r[n] \circ U_c h[n-1] + b_c)
\end{equation}
\begin{equation}
    h[n] = z[n] \circ h[n-1] + (1 - z[n]) \circ c[n]
\end{equation}
Where $W_z,W_r,W_c$ are the kernel weights for each gate,
$U_z,U_r,U_c$ are the recurrent weights for each gate, and
$b_z,b_r,b_c$ are the biases for each gate. Note that as the
inputs and outputs to the GRU layer may be vectors, all products
in the above equations are assumed to be standard matrix-vector
products, except those Hadamard products denoted $\circ$.
$\sigma(x)$ refers to the sigmoid function
$\sigma(x) = \frac{1}{1 + e^{-x}}$.
\subsubsection{Fully Connected Layer}
A fully connected layer computes an output vector $y[n]$ from
input vector $x[n]$ as follows:
\begin{equation}
    y[n] = \alpha(W x[n] + b)
\end{equation}
Where $W$ is the kernel weights, $b$ is the layer bias, and $\alpha(x)$
is the layer activation. In our model, we use no activation, i.e.,
$\alpha(x) = x$.

\subsection{Training Data}
Our dataset consists of $\sim 4$ minutes of electric guitar recordings,
from a variety of electric guitars including a Fender Stratocaster
and a Gibson Les Paul. The guitars are recorded ``direct'' meaning
that the recorded signal is equivalent to the signal received by the
pedal coming directly from the guitar. Recordings were made using a
Focusrite Scarlett audio interface at 44.1 kHz. Note that this
sample rate is very low compared to that used for other neural network
models of nonlinear audio effects (e.g. \cite{WaveNetVA,VArnn}). This sample
rate was chosen because the embedded hardware on which the final model
was implemented processes audio at this sample rate. The recordings
were then separated into segments of 0.5 seconds each, resulting in a
total of 425 segments.
\newline\newline
Since the original Klon Centaur pedal is quite expensive ($> 1500$ USD),
we used a SPICE simulation of the Centaur circuit in order to obtain a
``ground truth'' reference dataset. The reference dataset measures the
output voltage of the summing amplifier from the circuit at five
different values for the ``Gain'' potentiometer.
\begin{figure}
    \centering
    \includegraphics[width=0.5\textwidth]{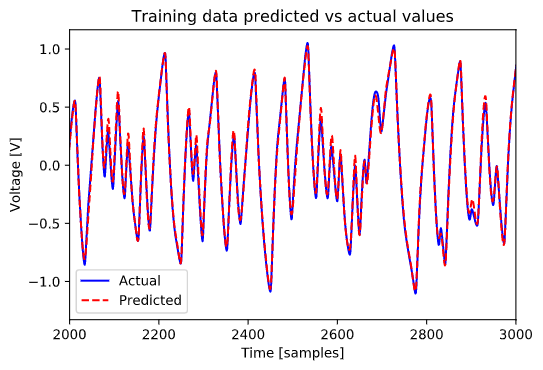}
    \centering
    \includegraphics[width=0.5\textwidth]{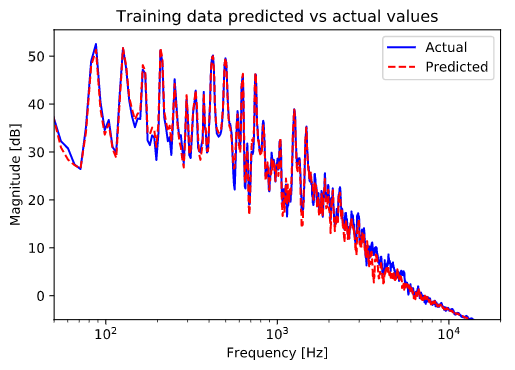}
    \caption{\label{fig:freqandtime} {\it Comparison of
        predicted output of the model against reference
        output shown in the time domain (above) and frequency
        domain (below). The frequency domain plot uses frequency
        band smoothing using $1/24$ octave bands for improved clarity.}}
\end{figure}
\begin{figure}
    \centering
    \includegraphics[width=0.5\textwidth]{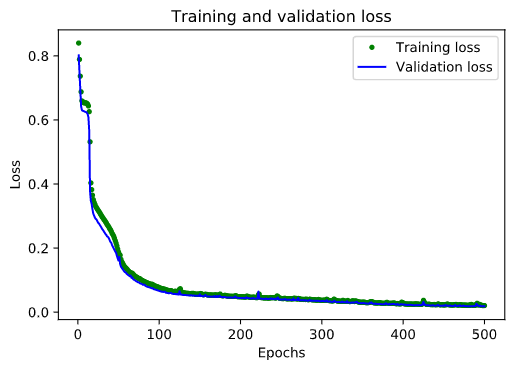}
    \caption{\label{fig:training} {\it Training accuracy
        for the RNN, shown over epochs.}}
\end{figure}

\subsection{Training}
We trained our models on 400 of the 425 audio samples, saving 25
samples for validation. Training was performed using the Adam
optimizer \cite{Kingma2015AdamAM}, with an initial learning rate
of \num{2e-3}. Each model was trained for 500 epochs; each training
session ran for $\sim 8$ hours. Similar to \cite{VArnn}, we use an
error-to-signal ratio (ESR) as the loss function for our models. For
a signal of lenght $N$, ESR is defined as:
\begin{equation}
    \mathcal{E}_{ESR} = \frac{\sum_{n=0}^{N-1} |y[n] - \hat{y}[n]|^2}{\sum_{n=0}^{N-1} |y[n]|^2}
\end{equation}
where $y[n]$ is the reference output, and $\hat{y}[n]$ is the predicted
output of the network.

\subsubsection{Training Results}
For each model, the trained network achieved a validation ESR of less than
$2\%$. Training and validation accuracies are shown in \cref{table:train}.
The training accuracy over epochs is shown in \cref{fig:training}. Results
comparing the output of the network to the reference output are shown in
\cref{fig:freqandtime}. Note that the high frequency response of the RNN
output is slightly damped compared to the reference.
\begin{table}[h!]
    \centering
     \begin{tabular}{||c | c | c||} 
     \hline
     Gain Parameter & Training ESR & Validation ESR \\
     \hline\hline
     0.0  & 0.50 & 0.70 \\
     0.25 & 0.51 & 0.57 \\
     0.5  & 0.57 & 0.50 \\
     0.75 & 0.70 & 0.67 \\
     1.0  & 1.63 & 1.72 \\
     \hline
     \end{tabular}
    \caption{\label{table:train} {\it Training and validation
        accuracies given in error-to-signal ratio percentages
        for each trained RNN model.}}
\end{table}

\subsection{Advantages and Limitations}
The recurrent neural network is a flexible and powerful
black-box modelling tool for stateful nonlinear systems.
The main limitation of the RNN model is its computational
complexity for large models, mostly due to the fact that the
$\tanh$ and sigmoid functions required by the recurrent layer
can be costly to compute. Further, it can be difficult to include
control parameters in the model, a persistent challenge with
black-box approaches. Finally, the recurrent neural network
cannot be used at arbirtary sample rates, and must be trained
at the same sample rate that is used for processing.

\section{Implementation}
\begin{figure*}
    \centering
    \includegraphics[width=1.0\textwidth]{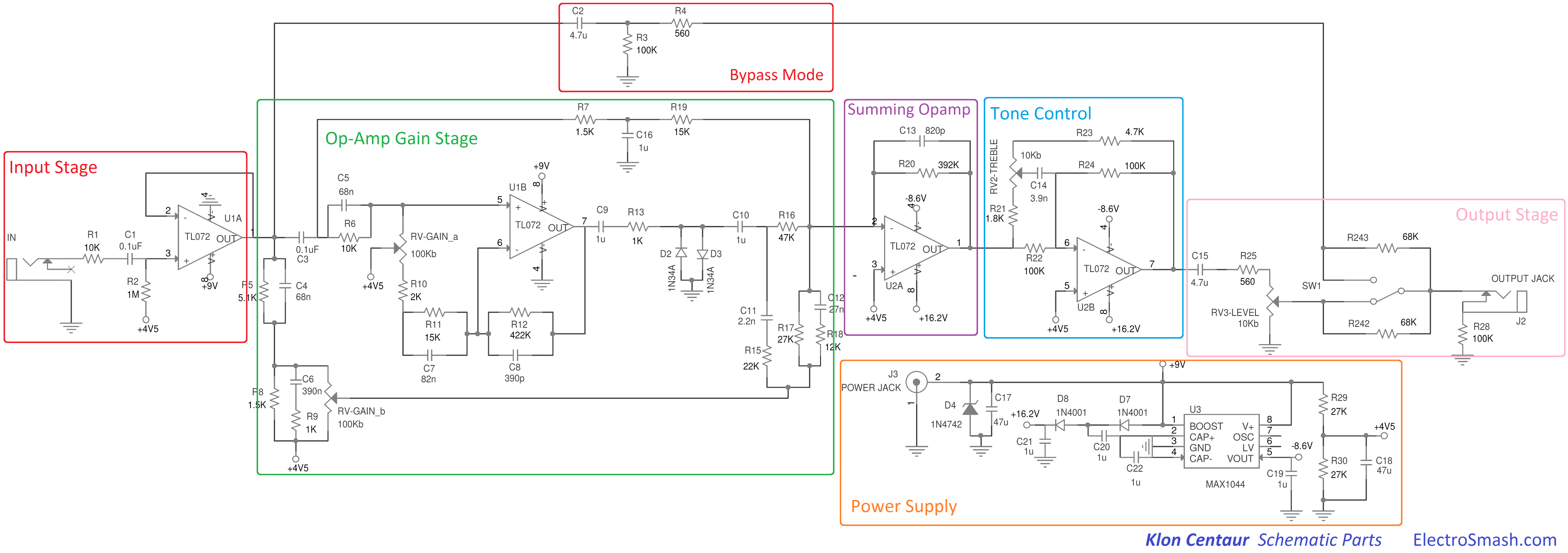}
    \caption{\label{fig:fullcircuit} {\it Full circuit schematic
    for the Klon Centaur guitar pedal with different circuit
    sections outlined. Adapted from \cite{electrosmash}.}}
\end{figure*}

\begin{figure}
    \centering
    \includegraphics[width=0.5\textwidth]{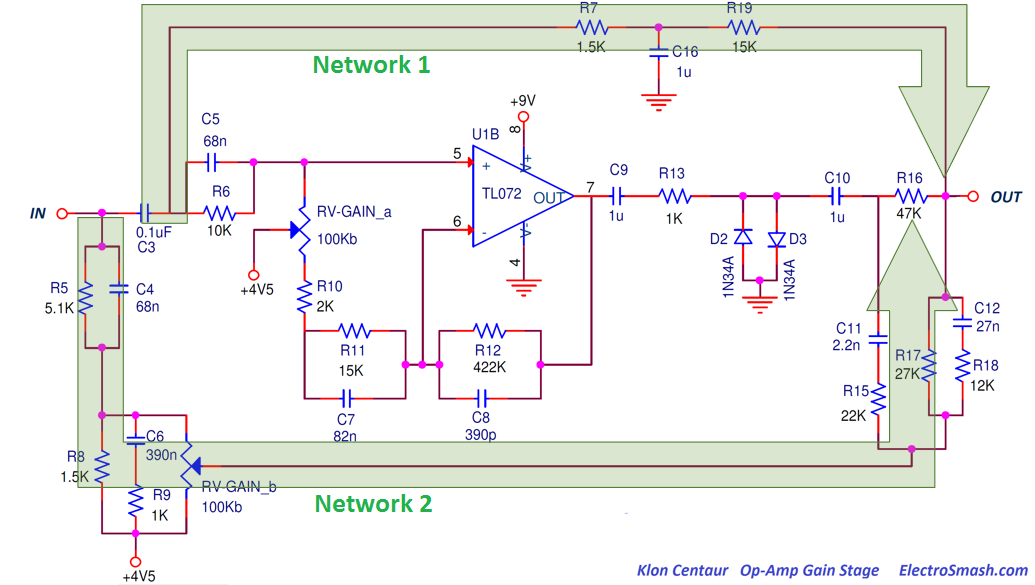}
    \caption{\label{fig:gaincircuit} {\it Circuit schematic for the
    gain stage from the Klon Centaur pedal, with the two Feed-Forward
    networks highlighted. Adapted from \cite{electrosmash}.}}
\end{figure}

In order to compare the virtual analog methods described above,
we construct two emulations of the Klon Centaur circuit: one emulation
using traditional circuit modelling methods (non-ML implementation), and
a second using a recurrent neural network (ML implementation). The Centaur
circuit can be broken down into four separable parts (see \cref{fig:fullcircuit}):
\begin{enumerate}
    \item Input Buffer
    \item Gain Stage
    \item Tone Control
    \item Output Buffer
\end{enumerate}
Due to their relative simplicity and linearity, in both emulations the
input buffer, output buffer, and tone control circuits were modelled using
nodal analysis. The ``Gain Stage'' circuit can be further broken down
into six (mostly) separable parts (see \cref{fig:gaincircuit}):
\begin{enumerate}
    \item Feed-Forward Network 1 (FF-1)
    \item Feed-Forward Network 2 (FF-2)
    \item Pre-Amp Stage
    \item Amplifier Stage
    \item Clipping Stage
    \item Summing Amplifier
\end{enumerate}
In the ML implementation, we treat the Gain Stage as a black box
with a single user-facing control (the ``Gain'' control). The RNN
model is designed to completely replace the Gain Stage in the circuit
model. In the non-ML implementation, we use nodal analysis to
model the amplifier stage, and summing amplifier circuits.
For FF-2 and the clipping stage, we use a wave digital filter.
Since FF-1 and the pre-amp circuit share a capacitor, we construct
a joint WDF model of these two circuits, using the voltage output
from the pre-amp circuit as the input to the amplifier stage, and
the current output from FF-1 (summed with the current outputs of
FF-2 and the clipping stage) as the input to the summing amplifier.
\begin{figure}
    \centering
    \includegraphics[width=0.5\textwidth]{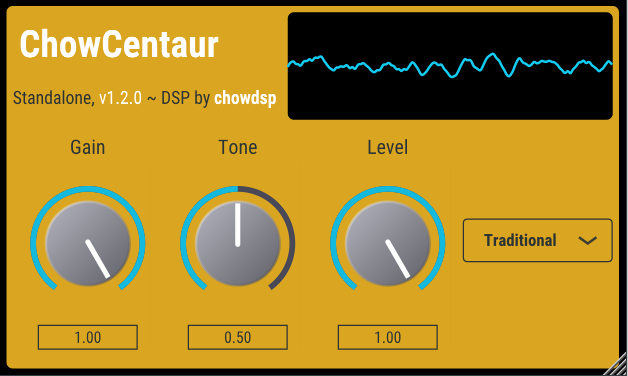}
    \caption{\label{fig:Plugin} {\it Audio plugin implementation
    of the Klon Centaur circuit model. Note controls for ``Gain'',
    ``Tone'', and ``Level'' analogous to the original circuit,
    as well as the ``Traditional/Neural'' parameter to control whether
    the emulation uses the traditional circuit model, or the RNN model.}}
\end{figure}

\subsection{Audio Plugin}
Digital audio effects are often implemented as audio plugins that
can be used by mixing engineers, producers, and musicians in a
consumer digital audio workstation (DAW) software. Common plugin
formats include the Avid Audio Extension (AAX), Steinberg's Virtual
Studio Technology (VST), and Apple's Audio Unit (AU) for desktop use,
as well as Apple's Audio Unit v3 (AUv3) for mobile use. The JUCE C++
framework\footnote{\url{https://github.com/juce-framework/JUCE}} is
commonly used to create cross-platform, cross-format plugins.
\newline\newline
As a demonstration of the two circuit emulations, we construct an audio
plugin containing both models, allowing the user to switch
between the two models for comparison. The plugin is implemented using
JUCE/C++, along with a real-time Wave Digital Filter
library\footnote{\url{https://github.com/jatinchowdhury18/WaveDigitalFilters}}
for the WDF models. For computing the output of the RNN models, we have
implemented a custom inferencing engine in C++, with two modes, one using
the Eigen linear algebra library \cite{eigenweb}, the second using only
the C++ standard library. In the future, we plan to add a third mode that
uses the Tensorflow Lite library.\footnote{\url{https://www.tensorflow.org/lite/}}

\subsection{Embedded Implementation}
Digital audio effects are sometimes implemented on embedded devices
for use in stage performances, often in the form of a guitar pedal,
or synthesizer module. Deploying an audio effect on an embedded device
can be difficult, due to the constraints in processing power and memory
availability. Further, in order to achieve a more expressive performance,
musicians often prefer effects that add minimal latency to the signal,
meaning that the embedded implementation must be able to run with a
very small buffer size.
\newline\newline
We chose the Teensy 4.0 microcontroller as our embedded platform, since
it contains a reasonably powerful floating point processor at a relatively
low price point. The Teensy can be purchased along with an Audio Shield,
which provides 16-bit stereo audio input/output at 44.1 kHz sampling rate.
The Teensy has gained popularity in the audio community due to
the Teensy Audio Library\footnote{\url{https://www.pjrc.com/teensy/td_libs_Audio.html}}
that contains useful audio DSP functionality, as well as the Faust
programming language which allows audio effects and synthesizers made in
Faust to be exported for use on the Teensy \cite{Michon2019RealTA}.
The Teensy 4.0 with the audio shield can be purchased for 35 USD.
\newline\newline
The Teensy implementation is writteen in C++ using the Teensy Audio Library,
along with the same WDF library as the audio plugin, and the standard
library mode of the same RNN inferencing engine. The emulation can be compiled
to use either the ML or non-ML implementation. Variables in the code can be
connected to potentiometers or push-buttons to control model parameters in real-time.
\begin{figure}
    \centering
    \includegraphics[width=0.5\textwidth]{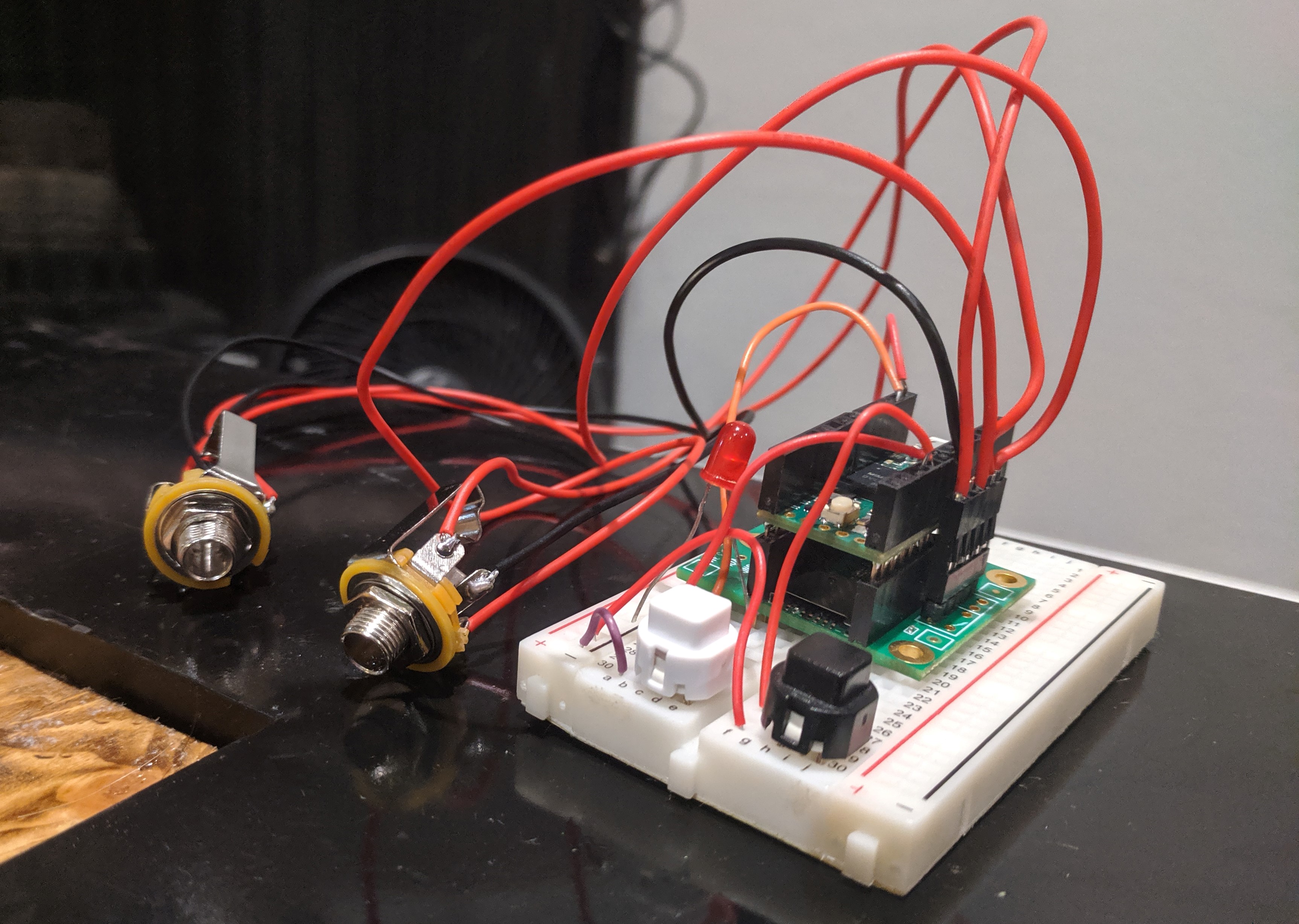}
    \caption{\label{fig:Teensy} {\it Teensy microcontroller implementation.}}
\end{figure}

\section{Results}
The results of the real-time implementations described above can best
be seen through audio performance examples. To that end, we provide
video examples of both implementations being used in real-time on a
guitar input being performed live. These examples can be seen on
YouTube.\footnote{\url{https://www.youtube.com/playlist?list=PLrcXtWXbPsj11cNBamVyMmDcWY1SXZHvz}}
From subjective listening, the ML and non-ML implementations sound
very similar, although the ML implementation has slightly damped
high frequencies, as predicted by the results of model training
(see \cref{fig:freqandtime}). The high frequency damping is slightly
more noticeable when the audio input is something other than a
guitar, e.g. drums. These issues could likely be alleviated
by training on a more diverse set of audio, and possibly by adjusting
the loss function to weight high frequencies more.

\subsection{Performance}
We also evaluate the computational performance of the emulations.
For real-time performance it is important to have fast computational
performance in order to reduce audio latency. In \cref{table:bench},
we show the compute time per second of audio processed of the various
models at different input block sizes. Note that at all block sizes,
the ML implementation outperforms the non-ML implementation. Performance
evaluation was completed using a 2017 Dell Precision laptop with a 2.9
GHz Intel Core i7 processor.
\begin{table}[h!]
    \centering
     \begin{tabular}{||c | c | c||} 
     \hline
     Block Size & NonML Time & ML Time \\
     \hline\hline
     8    & 0.0723437 & 0.0528792 \\
     16   & 0.0703079 & 0.0510437 \\
     32   & 0.0652856 & 0.0511147 \\
     64   & 0.0662835 & 0.0502434 \\
     128  & 0.0666593 & 0.0495194 \\
     256  & 0.0696844 & 0.0480298 \\
     512  & 0.0669037 & 0.0477946 \\
     1024 & 0.060816  & 0.0488841 \\
     2048 & 0.0695175 & 0.0488309 \\
     4096 & 0.0623839 & 0.0472191 \\
     \hline
     \end{tabular}
    \caption{\label{table:bench} {\it Benchmark results
        comparing processing speed of the audio plugin
        implementation using ML processing vs. non-ML
        processing. Speed is measured in compute time per
        second of audio processed.}}
\end{table}

\subsection{Recommendations}
From the process of implementing the circuit emulations described
above, we provide the following recommendations for circuit modellers:
\begin{itemize}
    \item For simple, linear circuits, nodal analysis is the easiest
          and most performant circuit modelling method.
    \item When modularity is important, prefer Wave Digital Filters.
          This modularity can refer to the circuit topology, the
          components in the circuit, or the way in which the components
          are discretized.
    \item For complex nonlinear systems, particularly systems with
          multiple nonlinear elements, or stateful nonlinear topologies,
          consider using recurrent neural networks.
    \item Small RNNs can outperform more complex circuit modelling methods
          while still maintaining model accuracy.
    \item While handling control parameters with RNNs can be difficult,	
          this can be acceptably solved by training multiple models for	
          different values of the control parameter and fading between them	
          in real time.
\end{itemize}

\section{Conclusion}
We have constructed two emulations of the Klon
Centaur guitar pedal circuit, using circuit modelling techniques
including nodal analysis, wave digital filters, and recurrent
neural networks. We described and compared the advantages and
limitations of each method, and showed how they can be used together
to achieve good results. We implemented the circuit emulations in the
form of an audio plugin and guitar-pedal style effect embedded on
a Teensy microcontroller. Finally, we provided recommendations for
utilising different circuit modelling methods for different types
of circuits, and for different platforms. The code for both
implementations, as well as the model training, is open source
and can be found on GitHub.\footnote{\url{https://github.com/jatinchowdhury18/KlonCentaur}}
\newline\newline
In future works, we would like to extend the RNN framework to be able
to implement larger networks in real-time. Specifically, the Differentiable
Digital Signal Processingg (DDSP) library from Google's Magenta project
implements complex audio effects including timbral transfer, dereverberation,
and more, using an auto-encoder that contains two 512-unit GRUs, along with
several other complex operations \cite{engel2020ddsp}. Being able to
implement the DDSP auto-encoder for use on real-time signals
would be a powerful tool for musicians and audio engineers.

\section{Acknowledgments}
The author would like to thank Pete Warden and the EE292 class at
Stanford University for inspiring this project, as well as Julius
Smith, Kurt Werner, and Jingjie Zhang for assistance with Wave
Digital Filter modelling. Thanks as well to the Center for Computer
Research in Music and Acoustics (CCRMA) for providing computing
resources.

\nocite{*}
\bibliographystyle{IEEEbib}
\bibliography{references}

\end{document}